\documentclass[12pt]{article}
\usepackage{amsmath,amssymb}
\begin{document}

\begin{center}
 {\Large  Synthesis of superheavy elements and theory of
atomic nucleus\\ }
\vspace*{2cm}
 {\large  B.N.Kalinkin , F.A.Gareev\\}
\vspace*{1cm}
 {\large BLTPH JINR, Dubna,\\ Russia \\}
\end{center}
\vspace*{5cm}
\begin{abstract}

The connection of synthesis of  long-lived superheavy nuclei and
existence of the island of stability with the choice  of  
realistic nuclear potential is discussed: the behaviour of the
average field of nuclei is an important element of nuclear models.

It is also shown that it is just Dubna that possesses the priority
both in the recent synthesis of a superheavy nucleus with charge
$Z=114$  and in its theoretical prediction made 35 years ago.
Possible sizes of the island of stability of superheavy nuclei are
considered.
\end{abstract}

\newpage

\section{Introduction}
\begin{sloppypar}

The synthesis recently performed for a superheavy long-lived nucleus (SHN)
with the charge $Z=114$ \cite{Og99} that is in a beyond transuranium region of mass
numbers was not a great surprise. The possibility of its existence was
theoretically predicted as early as in 1966 \cite{Gar}, i.e. long before experimental
synthesis. Nevertheless, this event was met with great interest and produces
a lively discussion. Recently, there appear papers \cite{Og20} and reviews
\cite{Her,HO20} devoted to that regions of studies.

Unfortunately, in our opinion, some of that admit considerable inaccuracies,
errors, and even distortions of the known historical  facts. Their serious
drawback is also the descriptive aspect that dominates the studies, but there
is completely absent the accent on the important problem of significance of
the experimental synthesis of SHN for the theory of atomic nuclei. But this
should be the main purpose of experiment.

In this paper, we briefly formulate our point of view on the problem as a
whole and its significance for the construction of nuclear theory

\section{Essence of the theoretical aspect of the problem}

The problem is directly connected with the experimental fact: nuclei with
$Z, N=8,\;20,\;28,\;50,\;82$
(and for neutrons with $N =126$) are the most stable to various decay modes.
Interpretation of this phenomenon became possible where there appeared the
shell model \cite{Nem,Pres}, according to which "magic" numbers are occupation numbers
of one-particle levels in nuclei, upon which the spectrum acquires a
considerable energy gap, and the binding energy becomes the highest.
Consequently the theoretically predicted existence of superheavy nuclei beyond
the limits of the periodic Table should at least be based on the calculation
of one-particle proton and neutron spectra for detecting considerable energy
gaps in them. Then, on the basis of the structure of levels of the found new
shell, one should estimate the SHN stability to the most probable decay mode--
fission, i.e. determine the barrier $V_{f}$ for it.

Of course. the problem can immediately be "closed " if it is assumed that the
next to the $Z=82$ proton closed shell is the $Z=126$ shell. And for the long time,
this was the opinion of most physicists.

It was no mere chance that parameters of the widely known Nilsson scheme \cite{Nils}
constructed on the oscillatory potential were more often fitted so as to
reproduce the shell $Z=126$ in the region of large $A$, i.e. the prediction was in
fact substituted by the assumption of the Nilsson scheme being valid for the
description of a system of SHN levels.
However, as early as in the mid-sixties when interest in the [problem revived,
some theoreticians understood that the oscillatory potential (the Nilsson
scheme \cite{Nils}) is not valid for this purpose. We think, the only advantage of it
is the simplicity of wave functions of one-particle states. From a physical
point of view, the largest drawback of it leading to a number of consequences
is that it tends to infinity near the nuclear surface. As a result, the wave
functions of the one-particle spectrum exhibit a wrong behavior at the
boundary and periphery of a nucleus---in the region that decisively
contributes to the probabilities of radiative transitions (the transition
operator $r^{\lambda}Y_{\lambda,\mu}(\theta,\phi)$, $\lambda=1,2,3,...$)
as well as of elastic, inelastic scattering, and other
reactions. Therefore, large discrepancies are inevitable (sometimes, by an
order of magnitude) with calculations on "correct" wave functions.

A serious drawback of that scheme is the necessity of changing the
parameters of the potential and spin--orbital interaction when passing to a
higher shell. Therefore, it is not surprising that the calculations of the
spectrum in the region of superheavy nuclei based on the distant extrapolation
of the parameters of the potential and spin--orbital interaction led to
different magic numbers for Z and N. For instance, Nilsson et al. \cite{Nils}
obtained $Z=126$, $N=164, 184$. There arose other solutions, as well. It is
evident that this
scheme cannot be recognized convinced and reliable, especially, for
predictions.

An acceptable solution to the problem should use the realistic potential $V(r)$
as the mean nuclear field. The term "realistic" means in the case that:
1)The potential $V(r)$ should be finite in magnitude (this follows, for
instance, from finiteness of such nuclear characteristics as the binding
energy of a nucleon and the limiting Fermi energy in the ground state of
nuclei;
2)near $r=R_{0}$ ($r=R_{0}$ is the radius of a nucleus), $.$ should be smeared owing to
quantum fluctuations of the nucleon density (especially, weakly coupled
nucleons) in the surface layer of a nucleus;
3) it is preferable to use  the spin--orbital interaction
$V_{SO}(r)$  its most justified form proposed by Heisenberg in analogy with the
electromagnetic field \cite{Geiz}

$$V_{SO}=\kappa(\vec{l}\bullet\vec{S})\frac{1}{r}\frac{dV_{N}}{dr},\eqno(1)$$

4) the electric charge $Ze$ should be distributed over the nucleus. In the
first approximation for the proton--nucleus Coulomb interaction, one can
assume

$$V_{c}(r)=\frac{(Z-1)e^{2}}{r}[\frac{3}{2}\frac{r}{R_{0}}-
\frac{1}{2}(\frac{r}{R_{0}})^{3}];\;r\geq R_{0}$$
$$V_{c}(r)=\frac{(Z-1)e^{2}}{r};\; r\leq R_{0}. \eqno(2)$$

The most convenient form of the potential $.$ in the case of spherical nuclei
(magic and near-magic ones) was suggested by Saxon and Wood \cite{Sax}:

$$V(r)=-V_{0}(1+exp\frac{r-R_{0}}{a})^{-1},\;\;R_{0}=r_{0}A^{1/3}.\eqno(3)$$

Studies performed in the framework of the shell model and scattering
theory  \cite{Nem} on nuclei have shown that expressions (1) and (3) should be
improved by taken account of the dependence of parameters
$V_{0}$ and $\kappa$ on $N$ and $Z$
as follows:

$$V^{N,Z}(r)=-V^{N,Z}_{0}(1+exp\frac{r-R_{0}}{a})^{-1},\;\;
R_{0}=r_{0}A^{1/3},\eqno(4)$$
where
$$V^{N}=V_{0}[1+0.63\frac{N-Z}{A}],\; V^{Z}=V_{0}[1-0.63\frac{N-Z}{A}],$$
$$V_{0}=53\; MeV,\;R_{0}=r_{0}A^{1/3};
\;r_{0}=1.27\;fm,\;a=0.67\; fm, \eqno(5)$$
and in formula (1) $\kappa$ should be taken in the form

$$\kappa=0.263(1+\frac{N-Z}{A}).\eqno(6)$$

However, calculations of one-particle energies and wave functions with that
potential even for spherical nuclei can be carried out only numerically (see,
i.e. \cite{Nem66}). A basic drawback of these calculations is that the wave
functions are obtained in the form of cumbersome tables; and their use in
numerous applications is rather inconvenient (this assertion was absolutely
correct  30 years ago, but now, the situation has changed).

In ref. \cite{Kal}: Kalinkin, Grabovskii, and Gareev "On levels of mean field of
nuclei" (JINR Preprint P-2682; accepted for publication in Acta Physica
Polonica on May 23, 1966), we developed an original method of a semianalytic
solution to the Schroedinger equation with the realistic potential. According
to this method, the radial wave functions can be represented in the analytic
form
$$\psi_{n}(r)\sim constH_{n}(S)exp(-S^{2}/2),\eqno(7)$$
where $n$ are radial wave numbers, $H_{n}$ are Hermite polynomials,
and $ S(r) $
is a correcting function of the form

$$
S(x)=\left\{\begin{array}{ll}
b_{1}ln(r/a_{0}) &\mbox{if $r\leq a_{0}$}\\
b_{2}ln(r/a_{0}) &\mbox{if $r\geq a_{0}$}
           \end{array}
           \right.
$$
The constants $a_{0}$, $b_{1}$ and $b_{2}$  are determined numerically with the use of a simple
procedure (for details, see \cite{Kal}).

As it is to be expected, the state wave functions for the realistic potentials
differ considerable, and in some cases significantly, from the Nilsson ones,
which immediately influences the behavior of many theoretically calculated
nuclear characteristics (examples can be found in \cite{Gra}). Discrepancies are
sometimes of an order of magnitude.

Thus, it is completely clear that the prediction of existence of new closed
shells in the SHN region  can be based only on the study of one-particle
spectra constructed on the realistic potential.
In this case, to avoid arbitrariness, it is of fundamental importance to base
final results on the values of parameters $V_{0}$, $r_{0}$, $a$ and$\kappa$
fixed from the analysis
of data on independent processes such as excitation of low-lying states of
near-magic nuclei, reactions of one-nucleon transfers, processes of elastic
and inelastic scattering of nucleons on nuclei, including effects of
polarization.

Here it should immediately be stressed that this problem is a particular
element in the solution of a more general problem of further verification of
the characteristics of the nuclear potential describing averaged
"nucleon--nucleus" interactions. The nuclear potential plays a central part
in the most of the theoretical models that describe the structure of nuclear
states, since it makes the basis one-particle proton and
neutron states (energy levels and wave functions) are constructed. Transition
to the observable states is accomplished by taking account of the corrections
due to the action of residual correlation forces between nucleons (leading,
for instance, to the superfluidity effect).

Directions of further verification of the accepted form of the realistic
potential are quite clear. One should study its "work" i. in the transition
to large values of the mass number A and ii. the possibility of its extension
for the transition from the spherical form to the deformed form of a nucleus.

In the first case, this approach corresponds to the study of the SHN
region---the prediction and search for new shells, the study of the
hypothetical island of stability.

In the second case, this transition is connected with the study of properties
of strongly deformed nuclei on the basis of a strongly deformed
realistic potential. They include the nuclei of rare-earth and transuranium
groups of elements. Note that a positive result in this case is also important
for the study of SHN, since it provides a reliable basis for estimations of
their fission barriers $V_{f}$ within the framework of shell-correction method
(Meyers and Swiatecki \cite{Mey66} formulated first this method and its most
consistent form, was developed by V.M Strutinsky \cite{Str66}). In the next two sections, we consider in more detail the
studies performed along these directions.

\section{The first theoretical predictions of new closed shells with $Z>82$, $N>126$}

Besides the reasons formulated above, this section is necessary because of
the problem of priority. The situation existing in literature confirms this
necessity \cite{Kal2001}.

We present a chronological list of the first theoretical works devoted to the
prediction of new shells on the basis of the realistic nuclear potential
appeared in the first two years of the corresponding studies. It seems that
two years are the most optimal period of time (at least, in those years) for
the cycle "study--publication in a journal of general use". Later
publications of other authors cannot be considered to be fully independent.
For the same purposes, we supply the first works with a more detailed
information on the dates when the work was received by a journal and when it
was published.

Thus, the first theoretical predictions \cite{Gar} in accordance with the chronology
of their publications with brief commentaries:

1966

1. Gareev F.A., Kalinkin B.N., Sobiczewski A. "Closed Shells for $Z>82$ and
$N>126$ in a Diffuse Potential Well", Preprint JINR -2793, Dubna, 1966,
published  16 June 1966 and was then distributed by N.I. Pyatov among
participants of the
Int. Symposium on Why and how should we investigate NUCLIDES FAR OFF
THE STABILITY LINE, Lysekyl, Sweden, August 21-27, 1966;\\
A. Sobiczewski, F.A. Gareev, B.N. Kalinkin, "Closed shells for
$Z>82$ and $N>126$ in a diffuse potential well",
Phys. Lett. V.22, No 4(1966)500, received 22 July 1966, published 1 September
1966 \cite{Gar}.

In the paper, on the basis of the method developed in our paper \cite{Kal}, the
proton and neutron energy levels were calculated in dependence on $A$ at
$Z>82$ and $N>126$ for the realistic potential (1-3).
The results show that possible magic numbers are $Z=114$ and $N=184$. The
calculations were performed with the values of parameters indicated in the
previous section. The solution turned out to be stable with respect to the
variations of parameters of the potential and spin-orbital interaction
connected with a possible inaccuracy in their determination. No noticeable
energy gap was found in the system of levels near $Z=126$. Also, a rough
estimate
was given for the fission barrier ($V_{f}\approx 10$ Mev) of this doubly
magic nucleus $Z=114,\;N=184$.
 To this
end, we used the recipe considered in \cite{Swi63}. The value
$V_{f}\approx$ 7-8 MeV was obtained
recently \cite{SMO95} on the basis of a more accurate method of shell correction
developed by Strutinsky \cite{STR67}.

This work was the first publication that contains a clear statement on the
possibility of a superheavy nucleus with $Z=114$ to exist and expound the method
of solution and its stability within the framework of the realistic potential
with justified values of parameters.

1967

V.A. Chepurnov  "Average field of neutrons and protons  shells with
$Z>82$ and $N>126$, YaF, 1967, {\bf 6}, 955 (received 26 February 1967) \cite{Chep}.

In this paper, on the basis of numerical solution to the Schroedinger
equation with potential (2) and spin-orbital potential (1) and with the value
of parameters practically the same as in \cite{Gar}, systems of proton and neutron
levels including cases with $Z>82$ and $N>126$ were obtained. It was confirmed that new
shells with $Z=114$ and $N=184$ we have established are realized. Like in
\cite{Gar} , no signs of
the shell with $Z=126$ to exist were found. This paper contains a reference
to \cite{Gar} and the author mentioned the agreement between
his results and ours \cite{Gar}.

In this connection, Chepurnov pointed to a rather instructive result derived
in ref. \cite{Wo66}: Wong tried to obtain the proton shell with $Z=126$ on the basis of
Wood--Saxon potential. He was a success but at what price! He should increase
the value of the smearing parameter $a$ by 35 percent, which contradicts its
value determined from scattering processes.

Further corroboration of our results was obtained in the following works in
1967:

3. H. Meldner, "Predictions of new magic regions and masses for super-heavy
nuclei from calculations with realistic shell model single particle
hamiltonians",
Proc. of the intern. symposium..., Lysekyl, Sweden, August 21-27, 1966.
Received 14 September 1966, published 18 October 1967, Ark. Fys.
36(1967)593 \cite{Meld67}.

It was reported that new magic numbers should be $Z=114$ and $N=184$, and at the end
of the report it was noted that this investigation was carried out
independently of our study \cite{Gar}.

4. V.M. Strutinsky, Yu.A. Muzychka, in Proc. of Int. Conf. on Heavy Ion
Physics , Dubna, 13-19 October 1966., received on 16 October 1966,
published in November 1967, p.51 \cite{Str67}.

5. A.M. Friedman, "Calculations on the production of the next closed shell
nucleus and other nuclei" (ibid \cite{Fri67}).

In both the reports, using the realistic potential, the author came to the
same conclusion---$Z=114$ and $N=184$ are the most pronounced magic numbers in the region of
superheavy nuclei.

Thus, the theory based on the realistic potential determined with the help of
data on nuclear processes involving spherically symmetric nuclei predicted a
nontrivial result: Instead of the earlier expected closed proton shell with
$Z=126$ next to $Z=82$, there should be the shell with $Z=114$.

Great significance of this conclusion is quite evident: the experiment on
synthesis of the 114th element can be realized considerably easier.
Experimenters obtained a well-grounded orienting for organization of the
experiment on synthesis of superheavy nuclei \cite{Kal68,Gra69}.

\section{Transition from spherically symmetric nuclei to deformed ones}

Further development of the conception of the nuclear potential describing the
nuclear average field proceeds through its generalization to the case of strongly
deformed nuclei.

Success along this line has practical consequences also for solution of the
particular problem of search and study of superheavy nuclei:

First, the degree of reliability of the prediction of new shell increases;

Second, the accuracy of estimations rises for the fission barriers of
hypothetical SHN;

Third, it makes, in prospect, possible to study the states of deformed SHN.

For the first time, the problem of determination of the one-particle spectrum
of deformed axial-symmetric nuclei with the realistic potential was solved by
Nemirovskii and Chepurnov \cite{Nem66}. They used the method of numerical integration
of a system of differential equations. However, these calculations were
complicated and required much computer time (we recall that we are talking
about the seventies), and therefore, their further development was stopped.

In this section, we consider another, though approximate, but more effective
method of the investigation of one-particle states of deformed nuclei, we
have elaborated in \cite{Gar}. It allowed us to represent the solutions in analytic
form, which simplifies their use in applications, and to carry out
required calculations more rapidly by several orders. Besides, it was
shown in \cite{Gra} that the energies and wave functions of one-particle states
situated near the Fermi surface almost coincide.

Analysis of the rotational bands observed in strongly deformed nuclei and the
study of intensities of transitions inside every rotational band and between
various bands show that the parity, total momentum of a nucleus and its
projection onto the intrinsic axis of symmetry are good quantum numbers. And
this is possible only in the case when a nucleus possesses axial symmetry.
Therefore, it is natural to proceed from the assumption that the dependence
of the radius of a deformed axially symmetric nucleus on deformation
parameters $\beta_{20}$ and $\beta_{40}$ and angle $\theta$
(counted from the symmetry axis) is of the form

$$R(\beta,\theta)=R_{0}(1+\beta_{0}+\beta_{20}Y_{20}(\theta)
+\beta_{40}Y_{40}(\theta), \eqno(8)$$
where $R_{0}$ is the mean radius of a spherical nucleus; $\beta_{0}$ is a constant
determined from the condition of nucleus volume being conserved;
$\beta_{20}$ and $\beta_{40}$ are
parameters of the quadrupole ($\lambda=2$) and hexadecapole ($\lambda=4$)
deformations. In the
main regions of deformed nuclei $150\leq A\leq 190$ and $230\leq A \leq 260$
the inequalities $\beta_{20}>0$ and $\beta_{40}\neq 0$
hold valid \cite{PA70}. Then, for the nuclear potential, we can write

$$V(\beta,\vec{r})=-\frac{V_{0}}{1+exp[(r-R(\beta,\theta))/a]}, \eqno(9)$$
(in what follows, for simplicity, we omit indices $N$ and $Z$ of the potential
and indices (20) and (40) of the deformation parameters $\beta_{20}$ and
$\beta_{40}$. For spin-orbital
interaction, the potential should be written in the invariant form

$$V_{SO}(\beta,\vec{r})=-\kappa[\vec{p},\vec{\sigma}]\bullet gradV(\beta,\vec{r}),
\eqno(10)$$
where $\vec{\sigma}$ is the spin of a nucleon; and $\vec{p}$, its momentum.
Formula (10)
transforms into formula (1) when $\beta\rightarrow 0$. For a proton system, it is necessary to
add the Coulomb term

$$V_{C}(\beta,\vec{r})=\frac{3}{4\pi}\frac{(Z-1)e^{2}}{R^{3}_{0}}
\int\frac{n(\beta,\vec{r'})d\vec{r'}}{\mid\vec{r}-\vec{r'}\mid}, \eqno(11)$$
where the distribution density of the charge in a nucleus $n(\beta,\vec{r'})$
equals

$$n(\beta,\vec{r'})=\frac{1}{1+exp[(r-R(\beta,\theta))/a]}.\eqno(12)$$

For the spin-orbital interaction (10), we have
$$\tilde{V}_{SO}(\beta,\vec{r})=V_{SO}(\beta,\vec{r})-V_{SO}(r)=
W_{1}+W_{2}+W_{3}, \eqno(13)$$
where
$$W_{1}=-\frac{\kappa}{r}\frac{\partial}{\partial r}\tilde{V}(\beta,\vec{r})
(p_{\theta}\sigma_{\phi}-\frac{1}{sin\theta}p_{\phi}\sigma_{\theta})$$
$$W_{2}=-\frac{\kappa}{r^{2}sin\theta}\frac{\partial}{\partial\theta}
\tilde{V}(\beta,\vec{r})p_{\phi}\sigma_{r},$$
$$W_{3}=-\frac{\kappa}{r}\frac{\partial}{\partial\theta}
\tilde{V}(\beta,\vec{r})p_{r}\sigma_{\phi},\eqno(14)$$
where $p_{\theta}$, $p_{\phi}$ ¨ $p_{r}$ are momentum operators in spherical
coordinates; $\sigma_{\theta}$, $\sigma_{\phi}$ and $\sigma_{r}$ are Pauli
matrices. It should be remembered that the operator
$\tilde{V}_{SO}(\beta,\vec{r})$ is Hermitian, but the
operators $W_{1}$, $W_{2}$ and $W_{3}$ are non-Hermitian, therefore, the neglect of any of them leads
to the wave functions being nonorthogonal.

The condition of nucleus volume conservation can be write in the form

$$\int n(\beta_{00},\beta_{20},\beta_{40},\vec{r})d\vec{r}=
4\pi\int n(\beta=0,r)r^{2}dr.\eqno(15)$$
The equation (15) for determination of $\beta_{0}$ was solved by successive
approximation method at the given values of average field parameters.

Denoting the sum of the nuclear and Coulomb potentials and spin-orbital
interaction by $V(\beta,\vec{r})$, we perform the identical transformation
of the
Schroedinger equation
$$[-\frac{\hbar^{2}}{2m}\Delta+V(\beta,\vec{r})-E]\Psi=
[-\frac{\hbar^{2}}{2m}\Delta+V(\beta=0,r)
+\tilde{V}(\beta,\vec{r})-E]\Psi=0, \eqno(16)$$
where
$$\tilde{V}(\beta,\vec{r})=V(\beta,\vec{r})-V(\beta=0,r).\eqno(17)$$

Next, let us expand $\tilde{V}(\beta,\vec{r})$ in the series over spherical
harmonics
$$\tilde{V}(\beta,\vec{r})=\sum_{\lambda}C_{\lambda}(\beta,r)
Y_{\lambda 0}(\theta), \eqno(18)$$
where the coefficients $C_{\lambda}(\beta,r)$ are determined numerically.

We will look for the solution to equation (11) as the superposition

$$\Psi_{\nu}(\vec{r})=\sum_{nlj}a_{nlj}^{\nu}\psi_{nlj}^{K}, \eqno(19)$$
where $\psi_{nlj}^{K}$ are wave eigenfunctions of the Schroedinger equation with the
spherically symmetric potential $V(\beta=0,r)$:

$$[-\frac{\hbar^{2}}{2m}\Delta+V(\beta=0,r)-E_{nlj}]\psi_{nlj}^{K}=0, \eqno(20)$$
with
$$\psi_{nlj}^{K}=R_{nlj}(r)\sum_{\mu=\pm 1/2}(lK-\mu1/2\mu\mid jK)
Y_{lK-\mu}\sigma_{\mu}, \eqno(21)$$
where $l$ is an eigenvalue of the orbital moment,
$\vec{j}=\vec{l}+1/2\vec{\sigma}$, $K\equiv j_{z}$, $n$ is a radial
quantum number; and $R_{nlj}(r)$ is a radial part of the total wave function of the
form (7).  Superposition (19) should contain terms with $l$ of the same
parity. Since the potential is axial-symmetric, formula (19) does not contain
summation over $K$, and the projection of the angular momentum is in this case
an integral of motion.

Substituting expression (19) into (16), multiplying by $(\psi_{n'l'j'}^{K})^{*}$
from the left, and
integrating, we obtain

$$(E_{n'l'j'}-E)a_{n'l'j'}^{\nu}+\sum_{nlj}a_{nlj}^{\nu}<\psi_{n'l'j'}^{K}\mid
\tilde{V}\mid\psi_{nlj}^{K}>=0.\eqno(22)$$
Solving the system of equations (22), we can determine the values of energy $E$
and coefficients $a_{nlj}^{\nu}$ of superposition (19), i.e. the wave functions of
states.

The expounded approximate method was used to construct the one-particle basis
\cite{Gar68} for calculating the characteristics of deformed nuclei in the regions
$150\leq A\leq 190$ and $230\leq A\leq 260.$  Later, in 1970, a complex of
programs \cite{Gar71,Gar73,Sch70} was worked
out for calculations of properties of deformed nuclei in the framework of the
Bogoliubov--Soloviev model \cite{Sol71,Gri74,Sol81,Sol89}; and one block of that complex
calculates one-particle basis states by our method. The complex of programs
became a universal tool for concrete computations of the very diverse
characteristics of deformed nuclei. Parametrization of the average field and
residual interactions turned out to be so successful and adequate to the
formulated problem that even 30 years later, this complex is still used at
JINR and other nuclear-physics centers, and the characteristics of nuclei
calculated on the basis of this model serves as basis for identification and
description of experimental data (see, for instance, \cite{Ah20}).

To date, more than a hundred of works is performed and published. The most
important results are presented in a series of monographs (see \cite{Sol71,Gri74,Sol81}).
The general conclusion is that the form accepted in the cycle of our studies
for the average field is justified also for the case of strongly deformed
nuclei.

\section{Realization of the SHN synthesis}

Experimental verification of predictions of the theory could be accomplished
only about 30 years later. Synthesis of the $_{114}289$ element was realized at Dubna
at the end of 1998. It was possible due to purposeful efforts on modification
of the accelerating and measuring techniques and the development of
experimental procedures. It is quite clear that the very fact of successful
synthesis of the 114th element in the reaction $^{48}Ca+^{244}Pu$ gives evidence in favor of
a great achievement of the experimental nuclear physics. It initiated a chain
of new results obtained during the last two years. The following nuclides:
$_{114}287,\;_{114}288,\;_{114}289$ and $_{116}292$ were synthesized.

Also, the results of experiments in Berkeley were published \cite{Nin99}. According
to these publications, it was possible to observe the chain of decays of the
118th element
$$^{293}118\rightarrow ^{289}116\rightarrow ^{285}114 \rightarrow ^{281}112
\rightarrow ^{277}110 \rightarrow ^{273}Hs \rightarrow ^{269}Sg,$$

synthesized in the reaction $^{86}Kr+^{208}Pb$.

Of course, it is necessary to carry out further verifications and
coordination of these results. However, now we can with high confidence
make two conclusions:

1. The prediction of theory \cite{Gar} based on the approbated realistic potential is
experimentally verified: the new proton closed shell with $Z=114$ does exist.

2. The superheavy nucleus 114 (including its isotopes) is not single. In its
vicinity, there are other SHN forming the whole region. This result was
qualitatively predicted by the theory \cite{Kal99}: if there
exists a new closed shell, there should exist the island of stability
embracing a certain number of superheavy nuclei around it.

As to the second conclusion, its qualitative character should be emphasized:
there is still no theoretically correct procedure of estimation of the SHN
lifetime. Therefore, real sizes of the island of stability can be estimated
only approximately from general considerations. We formulate them as follows
\cite{Kal99}: Concerning the possibility of existence of the island of stability
rather than a single superheavy nucleus, in our opinion, evidence comes from
the very fact of synthesis of the nucleus with magic number of protons $Z=114$
and nonmagic number of neutrons $N=175$. Indeed, from here, there should follow
stability both of the doubly magic nucleus with $Z=114$ and $N=184$ (the island
center) and of the nucleus with nonmagic number of protons $Z>114$ but with magic
number of neutrons $N=184$. Also, nuclei with values of Z and N around the
indicated combinations should possess by enhanced stability. Rapid
development of studies in this region makes us hope that the validity of this
assertion can be verified experimentally in the nearest future.

To conclude this section, we recall once more that the theoretical prediction
of the superheavy nucleus with $.$ to exist, formulated for the first time at
Dubna \cite{Gar} that allowed experimenters to organize its purposeful search,
for many years later was confirmed with its actual synthesis at Dubna, too
\cite{Og99}.

Unfortunately, this fact is either not reflected in some papers or is
distorted.

\section{Conclusion}

What is the resume of the presented discussion? It can briefly be
formulated as follows:

1. Conception of the average field used in the modern theory of
atomic nucleus and realized in a particular choice of the
realistic nuclear potential is further confirmed by the SHN
synthesis. The realistic potential works also in the region of
extremely large mass numbers $A$. Significance of this fact for
further development of theoretical models of the nucleus structure
is obvious: if a theory not only explains but also successfully
predicts, its reliability as an important element in the
construction of nuclear models grows considerably.

It is also evident that the choice of the solution method of the
Schroedinger equation with the realistic potential, either
semianalytic (as it was done in our works \cite{Kal}) or
completely numerical, is of no theoretical importance, since the
results coincide with a good accuracy.

2. Satisfactory description of states of the known strongly
deformed nuclei based on the generalized realistic potential makes
us hope in its effective use also for studying the states of
deformed SHN of the island of stability.

In this connection, it is hoped that the estimate of the fission
barrier on the basis of spectra constructed on the realistic
potential and of the procedure of shell correction to the drop
energy of a nucleus developed by Strutinsky will turn out to be
the most consistent and exact.

3. The above consideration shows that the theoretical analysis ,
which is based on the allowance made for important nuclear
physical characteristics , studied before , and which allowed one
to predict the existence of long--lived superheavy nuclei in the
vicinity of the nucleus with $Z=114$ , and the synthesis of this
nucleus and neighboring nuclides , which has recently been carried
out , represent a unique study that started $35$ years ago and has
recently been finished successfully.

The authors are grateful to D.V.Shirkov for attention and support
in the development of the problem at JINR.

\section{Appendix}

In section 3, we described real chronology of the first theoretical works
devoted to prediction of existence of new closed shells in nuclei with
$Z>82$ and $N>126$. As we indicated in the Introduction, some authors of recent
publications admit considerable inaccuracies, errors, and even distortions of
the known historical facts. Below we present only two examples.

G. Herrmann in the review paper "From Nuclear Fission to Superheavy Elements"
(\cite{Her}, p.14) wrote:

{\sl ...Heiner Meldner (Arkiv Fysik {\bf 36}(1967)593) showed that shell
closedes should be expected at proton number 114 and neutron number 184, not too
far from the then heaviest elements and, thus, perhaps within rich...}

It is difficult to imagine that the author of this review does not know about our
paper \cite{Gar} --- it is published in the well-known journal --- "Physics Letters".
Even more, there is reference to our paper in H. Meldner's
paper quoted by the author.  There are references to our publication even in textbooks
\cite{Sol71,Fra} (see details in \cite{Kal99}).

Another example is the review by S.Hofmann and G.Munzenberg
"Discovery of the heaviest elements" (Rev. Mod. Phys., 72, 3, July, 2000).
We quote only
one sentence, in which the authors discuss principal important questions
(page 734):

{\sl The prediction of magic numbers was less problematic than the calculation
of the stability of those doubly closed shell nuclei against fission...

A series of papers have based calculations of the location and properties of
superheavy  elements (SHE's) on the Strutinsky shell-correction method
(Myers and Swiatecki, 1966; Meldner, 1967; Nilsson et al., 1968;
Mosel and Greiner, 1969; Fiset and Nix, 1972; Randrup et al., 1976)...}

The first affirmation  rouses astonishment. The reliability of
predictions of shell effects in the spectrum of one-particle nuclear states and
the origin of the energy gap depend on the substantiation of the choice of the
average nuclear field  and  on its values of  parameters to be fixed
established by the analysis of independent nuclear reactions (a wide range of
direct nuclear reactions, elastic and inelastic scattering of particles on
nuclei, the data from nuclear spectroscopy). It is well-known that
localization of new shells essentially depends on the form and parameters of a
nuclear potential.
     As for  the stability of heavy nuclei, both those existing  already  and
those to be predicted, it mainly depends on
the behaviour of "a shell correction" part to  "a drop" part of nuclear energy
depending on its deformation (according to Meyers and Swiatecki; Strutinsky).
The calculation of
"the shell correction" is by a concrete  shell scheme, and so, in spite of its
importance, it is secondary (it is clear from the term - "shell correction").

As it was shown in a preprint by B.N. Kalinkin and F.A. Gareev \cite{Kal99},
the first prediction
for  a new proton shell with Z=114 arising in state spectra for a realistic
potential with well-founded values of parameters and spin-orbital interaction
was formulated in papers by F.A. Gareev, B.N. Kalinkin, A.Sobiczewski,
Preprint JINR P-2793, June, 1966; A.Sobiczewski, Gareev F.A., B.N. Kalinkin,
Phys. Lett., 22, No.4, 1966, p.500.
The estimation of fission barrier  $V_{f}$ for a double magic
nucleus with z=114 and N=184  was also given in this work ($V_{f}=10$ MeV).
Unfortunately, the authors of
the review give reference to  this paper only for the discussion  of a
particular case --- the question about spin-orbital
interaction in nuclei removing it in background.

So, the second confirmation  including
reference to be contained in it completely twists the realistic situation and
 well-known facts.

First, Meyers and Swiatecki don't predict new magic
numbers. For demonstration, we quote fragment of their paper:

{\sl We do not wish to imply that there are grounds for believing
that any of these magic numbers (Z=126, N=184, 258) would show up
in practice, and we use them only to illustrate that some of the
consequences would be if a magic number turned out to be present
in the general neighborhood of super- heavy nuclei somewhat beyond
the end of the periodic table. The actual values of the magic
numbers might be different...

...In order to proceed in a realistic manner with discussion of
the existence and location of a possible island of stability beyond the
periodic table, the first requirement is the availability  of estimates for the
location and strength of magic number  effects in that region. When such
estimates have become available (through single - particle calculations in
realistic nuclear potentials), it will be possible to apply our semi-empirical
treatment of nuclear masses and deformations to the predictions of the fission
barriers of hypothetical super - heavy  nuclei.}

  We see, Myers and Swiatecki in their  calculations used the values of magic
numbers obtained by other authors with the  harmonic potential.
Estimates
on the basis of realistic potentials were not available for them  that time.

  Second , Myers and  Swiatecki  don't  use the Strutinsky-method as the
review
authors state. Strutinsky published his method in 1967, whereas
Meyers and Swiatecki developed their   method earlier(UCRL 1965 Nuclear
Phis.,1966). Moreover, the method (Meyers and Swiatecki) was severely
criticized by Strutinsky (\cite{STR66}, p. 526, see also \cite{STR67}, p. 420):

{\sl Recently Swiatecki has done an interesting attempt
of phenomenological description of shell effects and deformation in the nuclear
mass (W.D. Myers, W.J. Swiatechi. Nucl. Masses and Deformations,
UCRL-11980 (1965), W. Swiatechi. Proc. 2d Int. Conf. on Nuclidic Masses,
Spr.-Verlag,Viena, p. 58(1964)). Based on simplification of the  level model in deformed nuclei
and assumption of shell disappearance at some value of the deformation parameter
$\alpha_{0}$, an expression was proposed of the type
$S(N,Z)\sim exp[-(\alpha/\alpha_{0})^{2}]$ by him for describing the
correction to the "drop" mass of a nucleus. This method would be very
comfortable
but, unfortunately, a more complex expression is required for the shell effect
description. So ,it is easy to see that, in deformed nuclei in the Swiatecki
model,
we always have a positive mass correction and  the threshold of nuclei fission less
than the drop one and more strongly dependent on $Z^{2}/A$ than in the ordinary drop model.
Following our calculations, the shell correction $\delta V$ in the middle of the
shell for deformed nuclei is negative and is large in magnitude up to $\approx 3$ Mev.
So, because of the restriction of the expression for correction in the Swiatecki
model, practically,  we have the same energy  for flattened and stretched
states. And this is obviously wrong.}

    The third moment is that Meldner, despite the authors' desire,
doesn't consider the
dependence of the one-particle spectrum on deformation, and hence, in
his paper, he
did not use the Strutinsky method and,  moreover, did not estimate the
fission barrier
 for superheavy nuclei. So, he did not investigate the  problem of
their stability.

In accordance with arguments given above, we assume that there is the following
true formulation for the second confirmation: {\sl In a series of papers,
calculations were performed for the location and properties of newly predicted super heavy
elements (Sobiczewski, Gareev and Kalinkin, 1966; Chepurnov, 1967;
 Meldner, 1967; Nilsson et al., 1968;
Mosel and Greiner, 1969; Fiset and Nix, 1972; Randrup et al., 1976.}

  We emphasize once more that our paper was the first prediction of the existence
of nuclei in a new shell Z=114 and N=184 based on the  realistic potential and
 numerous data on nuclear reactions. The first estimation of the barrier
fission for this nucleus was also done. Its value is  close to recent
calculations \cite{SMO95}. A detailed
evidence of priority of this paper was given in \cite{Kal99}.

  It is a pity that the authors of the review devoted the whole chapter to the
problem of denomination of  new elements
 and did not pay any attention to principal questions.
\end{sloppypar}

\end{document}